\documentstyle[aps,preprint,floats]{revtex}

\tighten

\newcommand{\epem}{e^+e^-}

\newcommand{\susyU}{\widetilde{U}}
\newcommand{\dsusyU}{\widetilde{U}}

\begin{document}

\draft
\pagestyle{empty}

\preprint{
\noindent
\hfill
\begin{minipage}[t]{3in}
\begin{flushright}
FERMILAB--PUB--97/177--T \\
LBNL--40465 \\
UCB--PTH--97/33 \\
RU--97--45 \\
hep-ph/9706438 \\
June 1997
\end{flushright}
\end{minipage}
}

\title{Super-oblique corrections and non-decoupling \\
of supersymmetry breaking
}

\author{Hsin-Chia Cheng$^1$, Jonathan L. Feng$^2$
\thanks{Research Fellow, Miller Institute for Basic Research in 
Science.}, and Nir Polonsky$^3$}

\address{\vspace*{0.2in}
$^1$ Fermi National Accelerator Laboratory \\
P.~O.~Box 500, Batavia, Illinois 60510}

\address{\vspace*{0.07in}
$^2$ Theoretical Physics Group, Lawrence Berkeley
National Laboratory \\ 
and Department of Physics, University of
California, Berkeley, CA 94720}

\address{\vspace*{0.07in}
$^3$ Department of Physics and Astronomy \\ 
Rutgers University, Piscataway, NJ 08855-0849 } 

\maketitle

\begin{abstract}

If supersymmetric partners of the known particles have masses at the
multi-TeV scale, they will not be directly discovered at planned
future colliders and decouple from most observables.  However, such
superpartners also induce non-decoupling effects that break the
supersymmetric equivalence of gauge boson couplings $g_i$ and gaugino
couplings $h_i$ through supersymmetric analogues of the oblique
corrections.  Working within well-motivated theoretical frameworks, we
find that multi-TeV scale supersymmetric particles produce deviations
at the $1-10$\% level in the ratios $h_i/g_i$.  Such effects allow one
to bound the scale of kinematically inaccessible superpartners through
precision measurements of processes involving the accessible
superparticles. Alternatively, if all superpartners are found,
significant deviations imply the existence of highly split exotic
supermultiplets.

\end{abstract}

\pacs{11.30.Pb 14.80.Ly 11.10.Hi}

\pagestyle{plain}

\section{Introduction}
\label{sec:introduction}

Supersymmetric particles are often assumed to have mass on order of or
below the TeV scale if supersymmetry (SUSY) indeed plays a role in the
solution of the gauge hierarchy problem.  Otherwise, fine-tuning of
various parameters in the low-energy theory is
required~\cite{finetuning}, undermining the motivation for the
introduction of low-energy supersymmetry.  The prospects for
discovering and studying some supersymmetric particles (sparticles) at
present and future colliders is therefore promising, particularly at
the Large Hadron Collider (LHC) at CERN\cite{LHC} and proposed high
energy linear $\epem$ colliders\cite{JLC,NLC,DESYLC}.

It is, however, a logical possibility that only part of the sparticle
spectrum will be seen at planned future colliders, with some number of
superpartners of ordinary matter and gauge fields beyond the discovery
range.  In fact, such scenarios are realized in a wide variety of
models, and are often found in theories designed to solve the
supersymmetric flavor problem, {\em i.e.}, the problem that low-energy
constraints are violated for generic sfermion masses and mixings.
These models may be roughly divided into two categories.  In the first
class of models, which we will refer to as ``heavy QCD models,'' the
gluino and all the squarks are heavy.  Such may be the case in models
with gauge-mediated SUSY breaking\cite{gm}, where strongly-interacting
sparticles get large contributions to their masses, and also in the
no-scale limit of supergravity models\cite{noscale}.  These models
solve the SUSY flavor problem, since flavor-blind sfermion masses
result from the proportionality of sfermion masses to gauge couplings
and charges.  A similar spectrum may also be predicted in other
models, for example, grand-unified models with non-unified gaugino
masses and heavy gluinos~\cite{ACM}, in which the gluino drives the
squark masses up through renormalization group evolution.  In a second
class of models, the first and second generation squarks and sleptons
are very heavy with masses ${\cal O}(10 \text{ TeV})$, while the third
generation sfermions are at the weak
scale\cite{effectivesusy,u1,cohen,nilles}.  We will call these ``2--1
models.'' Such models are motivated by the desire to satisfy
low-energy constraints from, for example, $K^0 - \bar{K}^0$ mixing and
$\mu \to e\gamma$, without the need for sfermion universality,
sfermion alignment, or small $CP$-violating phases.  At the same time,
the extreme fine-tuning problem arising from very massive third
generation sfermions is alleviated. It should be noted, however, that
some increased level of fine-tuning must typically be tolerated, both
in these models\cite{effectivesusy,badfine} and in those of the first
category\cite{finetuning2}.
 
Given the many possibilities for supersymmetric particles beyond the
reach of the LHC and proposed $\epem$ colliders, it is well worth
considering what experimental implications such heavy states may have.
In most experimentally accessible processes, such states decouple, and
their effects rapidly decrease with increasing mass scale.  Here,
however, we study effects with the opposite behavior, that is, which
grow with increasing supersymmetric mass splittings.  Such effects
rely on the fact that the interactions in supersymmetric theories are
tightly constrained.  For example, SUSY implies the relations

\begin{equation}
\label{equivalence}
g_i=h_i \ , 
\end{equation}
where $g_i$ are the standard model gauge couplings, $h_i$ are their
supersymmetric analogues, the gaugino-fermion-sfermion couplings, and
the subscript $i=1,2,3$ refers to the U(1), SU(2), and SU(3) gauge
groups, respectively.  Unlike other relations, such as the unification
of gaugino masses, these relations hold in all supersymmetric models
and are true to all orders in the limit of unbroken SUSY.  However,
SUSY breaking mass differences within superfields with standard model
quantum numbers lead to corrections to Eq.~(\ref{equivalence}) that
grow logarithmically with the superpartner masses.  Deviations from
Eq.~(\ref{equivalence}) are thus unambiguous signals of SUSY breaking
mass splittings, and by precisely measuring such deviations in
processes involving accessible superparticles, bounds on the mass
scale of the kinematically inaccessible sparticles may be determined.

The corrections to Eq.~(\ref{equivalence}) are highly analogous to the
oblique corrections\cite{Peskin,oblique} of the standard model.  We
will therefore refer to them as ``super-oblique corrections'' and
parametrize them by ``super-oblique parameters,'' one for each gauge
group.  As is the case for oblique corrections, we will find that
super-oblique corrections are flavor-independent and are enhanced for
large heavy particle sectors.  Furthermore, the simple nature of the
corrections allows one to study them in a model-independent fashion
using only TeV-scale parameters.  As examples, we will calculate the
size of these corrections in the two classes of models described
above.  In both cases, we find substantial contributions to all three
super-oblique parameters.  Such corrections may be measured through a
variety of processes, depending on what sparticles are available for
study.  Tests of the SU(2) relation $g_2 = h_2$ with charginos have
been studied\cite{FMPT}, as has the possibility of testing and looking
for deviations in the U(1) relation with selectrons\cite{slepton}.
Soft SUSY breaking effects on hard supersymmetric relations, {\em
i.e.}, relations between dimensionless couplings such as
Eq.~(\ref{equivalence}), were also noted in Ref.~\cite{Hikasa}, where
such effects were calculated for the specific case of squark widths.
A general classification of possible observables at $e^{\pm} e^-$ and
hadron colliders, as well as detailed studies of representative
examples incorporating the variety of experimental uncertainties will be
presented in an accompanying study\cite{gex}.

We begin in Sec.~\ref{sec:decoupling} with a formal discussion of the
flavor-universal corrections to Eq.~(\ref{equivalence}).  The analogy
to the oblique corrections of the standard model is highlighted, and
super-oblique corrections and parameters are defined.  In
Sec.~\ref{sec:other} flavor-dependent corrections, as well as other
non-decoupling effects are discussed.  In Sec.~\ref{sec:framework} we
estimate the size of the super-oblique corrections in the heavy QCD
and 2--1 models described above.  The (typically small) contributions
of vector-like messenger and U(1)$'$ sectors to these deviations are
calculated in Sec.~\ref{sec:messenger}.  Our conclusions, as well as
additional comments concerning possible implications of measuring
super-oblique corrections, are collected in
Sec.~\ref{sec:conclusions}.

\section{Super-oblique corrections}
\label{sec:decoupling}

We would like to identify robust experimental signatures of
as-yet-undiscovered supersymmetric particles at future colliders.  If
only the standard model particles are available to us and we are only
able to probe momentum scales below sparticle thresholds, broadly
speaking, two approaches are possible.  The first is to look for their
virtual effects in low-energy processes. Unfortunately, in the models
discussed in Sec.~\ref{sec:introduction} with sparticle masses $\agt
{\cal O} (1-10\text{ TeV})$, such effects are often well below
experimental sensitivities.  This is just a statement of the
decoupling theorem\cite{AC} for heavy superpartners from low-energy
phenomena.

The second approach is to adopt some model dependent assumption such
that the values of the low-energy couplings may be interpreted as
signatures of heavy sparticles.  For example, if one assumes grand
unification boundary conditions for the gauge coupling constants,
their well-measured values at low energies are sensitive to sparticle
thresholds. Threshold corrections have been extensively studied both
with renormalization group techniques that incorporate leading
logarithm effects\cite{threshold1} and through explicit one-loop
calculations with finite corrections\cite{threshold2}.  Such
experimental signatures are, of course, model dependent and are
obscured by other effects, such as GUT scale threshold
effects.\footnote{Note, however, that it is possible that certain
processes probe momentum scales above sparticle thresholds, even
though no sparticles have been directly discovered.  By extrapolating
the low-energy couplings up to the characteristic momentum scales of
such processes, the presence or absence of intermediate sparticle
thresholds may be determined, independent of GUT assumptions.  The
possibility of such effects has been discussed, for example, in
Refs.~\cite{alphas,barger}.}

In this study, we will consider scenarios in which some, but not all,
superpartners are discovered.  As noted in
Sec.~\ref{sec:introduction}, such scenarios may be realized at future
colliders in a variety of models.  If this is the case, what may be
learned about the heavy, inaccessible superpartners?  It is well-known
that the decoupling theorem does not apply if the heavy particle
masses break symmetries\cite{AC}.  In the present case,
the heavy sparticle masses are predominantly invariant under standard
model symmetries.\footnote{Sfermion masses may break SU(2), but this
breaking is typically suppressed by the left-right mixing
$(m_{\text{fermion}}/ m_{\text{sfermion}})^2$.  Contributions of
sfermions to the SU(2) oblique parameters therefore may usually be
neglected\cite{susyrho}, and are especially small in the scenarios we
are considering, since the sfermion masses are at the multi-TeV
scale.}  However, these masses violate supersymmetry, and in fact, the
heavy superpartners give rise to non-decoupling corrections in
processes involving the light superpartners.  There are a variety of
non-decoupling effects that may be considered.  We will concentrate
here on a set which we will call ``super-oblique corrections,'' for
reasons detailed below.  These corrections are selected as
particularly important, because they are universal in processes
involving gauginos, enhanced by a number of factors, and may be
measured at colliders in a variety of ways.  Other non-decoupling
effects will be described in Sec.~\ref{sec:other}.

For simplicity, let us begin by neglecting the superpotential Yukawa
couplings and assuming both $R$-parity ($R_P$) conservation\cite{Rp}
and flavor conservation.  (The implications of relaxing these
assumptions are the topic of the following section.)  With these
assumptions, in processes involving standard model particles or the
light superpartners, the heavy superpartners appear at the one-loop
level only through renormalizations of gauge boson and gaugino
propagators.  These renormalizations are equivalent in the limit of
exact SUSY.  However, since the sparticles have SUSY breaking masses,
the corrections from the heavy sparticle loops are different for gauge
bosons and gauginos, and the effects are proportional to $\ln (M/m)$,
where $M$ ($m$) is the characteristic heavy (light) superpartner mass
scale.  These non-decoupling effects are similar in origin to the
logarithmically-divergent loop corrections to the Higgs boson mass in
supersymmetric theories\cite{higgs}.  In addition, they are process
independent, up to small ${\cal O}(p^2/M^2)$ corrections, where $p$ is
the momentum of the gauge bosons or gauginos, and can be absorbed into
the gauge couplings $g_i$ and gaugino couplings $h_i$.

It is instructive to draw an analogy between these effects and the
oblique corrections\cite{Peskin,oblique} of the electroweak sector of
the standard model.\footnote{This analogy was previously noted by
L.~Randall\cite{LR}.}  In the standard model, heavy particles with
isospin breaking masses enter low-energy observables dominantly
through the vacuum polarization functions of the electroweak gauge
bosons.  More specifically, SU(2) multiplets with custodial 
SU(2)\cite{custodial} breaking
masses, such as the $(t,b)$ multiplet, renormalize the propagators of
the $(W,Z)$ vector multiplet differently, leading to explicit
custodial SU(2) breaking in the vector multiplet at
the quantum level, and introducing non-decoupling effects that grow
with the mass splitting.  The supersymmetric non-decoupling
corrections may be described analogously with the following
replacements in the previous sentence:
\begin{itemize}
\item SU(2) multiplets $\to$ supermultiplets 
\item custodial SU(2) breaking masses $\to$ 
  soft supersymmetry breaking masses
\item $(t,b)$ multiplet $\to (\tilde{f}, f)$ supermultiplet
\item $(W,Z)$ vector multiplet $\to$ $(\text{gauge boson}, 
\text{gaugino})$ vector supermultiplet
\item custodial SU(2) $\to$ supersymmetry
\end{itemize}
Motivated by the strength of this analogy, we will refer to the SUSY
breaking effects of the heavy superparticles as ``super-oblique
corrections.''  As is the case for the oblique corrections of the
standard model, the super-oblique corrections provide a unique
opportunity to probe the scale of the heavy sector at low energies.

Let us investigate this analogy further.  The oblique corrections of
the standard model may be described in terms of the three parameters
$S$, $T$, and $U$\cite{Peskin}.  The latter two are measures of
custodial isospin breaking, with the differences of the mass and
wavefunction renormalizations of the $W$ and $Z$ (more correctly,
$W^3$) at $p^2=0$ from heavy particles given by $T$ and $U$,
respectively.  Below, we will define super-oblique parameters that are
measures of the splitting of $g_i$ and $h_i$.  Such coupling constant
splittings are results of differences in the wavefunction
renormalizations of gauge bosons and gauginos. The super-oblique
parameters we define are therefore most similar to $U$, and will be
denoted by $\susyU_i$, where the subscript $i$ denotes the
corresponding gauge group.

One might also hope that measureable supersymmetric analogues to $S$
and $T$ exist, especially since these are typically more sensitive
probes of new physics in the standard model.  The $S$ parameter is a
consequence of the extra U(1) gauge group, and is not a measure of
custodial SU(2) breaking.  There is therefore no analogous effect in
supersymmetry.  The analogue to $T$ is a difference in the mass
renormalizations of gauge bosons and gauginos.  In our case, there is
no mass renormalization of the gauge bosons due to the heavy
superpartners if their masses are standard model gauge
invariant.\footnote{By mass renormalization here we mean the mass
shift at $p^2=0$, {\em i.e.}, the part which is independent of
wavefunction renormalization. Note, however, that the physical masses
$m_W$ and $m_Z$ are renormalized by wavefunction renormalization.}  On
the other hand, gaugino masses may receive contributions from heavy
sparticle loops if these loops contain $R$-symmetry breaking effects.
If there were no tree level gaugino masses, or these masses were
somehow known, the loop-generated gaugino masses\cite{FaMa} 
would be a probe of
the mass splitting between components of a supermultiplet, providing a
probe analogous to $T$, parametrized by three new super-oblique
parameters $\widetilde{T}_i$.  However, in a general softly broken SUSY
theory, arbitrary gaugino masses already exist at tree level, and
there is no tree level mass relation between the gauge bosons and the
gauginos. (In contrast, custodial SU(2) symmetry enforces the relation
$m_W=m_Z\cos\theta_W$ at tree level in the standard model.)  The
gaugino mass renormalizations therefore may be absorbed into these
tree level terms, yielding no useful low-energy observables
corresponding to $T$, unless one makes some assumptions about the tree
level gaugino masses.

The non-decoupling SUSY breaking effects may also be profitably
understood in the language of renormalization group equations (RGE's).
Above the heavy superpartner scale $M$, SUSY is not broken, and we
have $h_i=g_i$.  Below $M$, where the heavy superpartners decouple,
light fermion loops still renormalize the gauge boson wavefunction
(and thus, $g_i$) but heavy sfermion loops and sfermion-fermion loops
decouple from gauge boson and gaugino wavefunction renormalization,
respectively. (Gauge loops still renormalize both wavefunctions in the
non-Abelian case.)  Since not all loops from the supermultiplet
decouple simultaneously, supersymmetry is broken in the gauge sector,
and therefore the gauge couplings $g_i$ and gaugino couplings $h_i$
start to evolve differently.

The one-loop evolution of the gauge couplings between the heavy and
the light superpartner scales gives

\begin{equation}
\frac{1}{g_{i}^{2}(m)}  \approx
\frac{1}{g_{i}^{2}(M)}  + \frac{b_{g_{i}}}{8\pi^{2}}\ln \frac{M}{m} \ ,
\label{g}
\end{equation}
where $b_{g_i}$ is the one-loop $\beta$-function coefficient of the
effective theory between the heavy and light mass scales, with the
heavy superpartners decoupled.  For the gaugino couplings, because
SUSY is broken, the RGE's will depend on both gauge couplings $g_i$
and gaugino couplings $h_i$.  However, because the deviations of $h_i$
from $g_i$ are small, the contributions from this difference to the RG
evolution are higher order effects and hence negligible.  In addition,
because $h_i \approx g_i$, the Ward and Slavnov-Taylor identities
still hold approximately for the gaugino couplings, and the primary
effect of the decoupled sparticles is to modify the one-loop
$\beta$-function coefficient of the gaugino coupling RGE.
Approximating $h_i \approx g_i$ in the RGE's, the gaugino couplings at
the scales of the light and heavy sectors are thus related by

\begin{equation}
\frac{1}{h_{i}^{2}(m)}  \approx
\frac{1}{h_{i}^{2}(M)}  + \frac{b_{h_{i}}}{8\pi^{2}}\ln \frac{M}{m} \ .
\label{h}
\end{equation}
The one-loop $\beta$-function coefficient $b_{h_i}$ is obtained by
subtracting the entire contribution of whole supermultiplets that
contain heavy superpartners.  Substituting
the supersymmetric boundary condition $g_{i}(M) = h_{i}(M)$,
straightforward manipulations yield

\begin{equation}
\frac{h_{i}(m)}{g_{i}(m)} \approx 1 + \frac{g_{i}^{2}(m)} 
{16\pi^{2}}(b_{g_{i}} - b_{h_{i}}) \ln\frac{M}{m} \ .
\label{hg}
\end{equation}

To parametrize the non-decoupling effects of heavy superpartners, we
define the super-oblique parameters

\begin{equation}
\susyU_{i} \equiv \frac{h_{i}(m)}{g_{i}(m)} - 1 
\approx \frac{g_{i}^{2}(m)}{16\pi^{2}}(b_{g_{i}} - b_{h_{i}})
\ln R \ ,
\label{delta}
\end{equation}
where $i=1,2,3$ denotes the gauge group, and $R = M/m$.  As noted
above, these parameters are supersymmetric analogues to the oblique
parameter $U$\cite{Peskin}, with one for each gauge group.  Note that,
because $b_{h_i} < b_{g_i}$, the coupling $h_i$ is more asymptotically
free than $g_i$, $h_i(m) > g_i(m)$, and the parameters $\susyU_i$ are
always positive.  This statement is true at the leading logarithm
level irrespective of whether the heavy sparticles are scalars or
fermions.  We may also define another set of parameters that are
deviations in the ratio of ratios, which we denote by the two-index
variables

\begin{eqnarray}
\dsusyU_{ij} &\equiv& \frac{h_{i}(m)/h_{j}(m)}{g_{i}(m)/g_{j}(m)} - 1
\approx \susyU_i - \susyU_j \nonumber \\
&\approx& \frac{1}{16\pi^{2}} \left[ g_{i}^{2}(m) (b_{g_{i}} -
b_{h_{i}}) - g_{j}^{2}(m)(b_{g_{j}} - b_{h_{j}}) \right] \ln R \ .
\label{rho}
\end{eqnarray}
These linear combinations of the super-oblique couplings are useful,
as they are probed by branching ratio measurements, which are
sensitive to $h_i/h_j$.

In fact, the decoupling scales for the gauge and gaugino couplings are
not identical when threshold corrections at the decoupling scale are
taken into account.  The finite threshold corrections slightly lower
the decoupling scales for the gaugino couplings relative to those of
the gauge couplings, which slightly reduces the deviations of $h_i$
from $g_i$ at low energy relative to the leading logarithm
analysis. However, these effects may be absorbed into an effective
heavy scale $M'$, with $R=M'/m$.  The finite corrections and the
resulting shift in $R$ are calculated in the Appendix.

\section{Other non-decoupling corrections}
\label{sec:other}

In the discussion above, we have examined a set of non-decoupling
corrections to the gaugino couplings that are universal in that they
apply to all gaugino couplings.  We have, however, neglected the
superpotential Yukawa couplings and have also assumed $R_P$ and flavor
conservation.  Such effects lead to additional non-decoupling
corrections, including flavor-specific gaugino coupling corrections.
In addition, couplings that do not involve gauginos also receive
corrections (even in the absence of Yukawa couplings and $R_P$ and
flavor violation).  Let us now consider each of these effects in turn.

In the presence of Yukawa couplings, new flavor-dependent
non-decoupling radiative contributions are possible.  For example, in
the minimal supersymmetric standard model, matter field wavefunctions
receive corrections from loops involving Higgs bosons and Higgsinos,
Higgs and Higgsino wavefunctions are corrected by loops involving
fermions and sfermions, and new contributions also appear in the
vertices.  These contributions grow logarithmically with the heavy
mass in the loop. Such effects spoil the approximate Ward and
Slavnov-Taylor identities for the gaugino couplings --- if a gaugino
coupling is renormalized by a Yukawa operator involving heavy
superpartners, the diagrams involving the heavy field decouple and the
cancellation of divergences is spoiled in the effective theory.  The
one-loop RGE's of the gaugino couplings will then also depend on
Yukawa couplings, and the universal gaugino coupling $h_i$ is split
into different couplings $h_i^f$ for each gaugino-$f$-$\tilde{f}$
vertex. These Yukawa coupling contributions to $\susyU_i$ are of the
opposite sign to the universal corrections discussed above. Of course,
such effects are typically suppressed by small Yukawa couplings and
are only relevant for processes involving the Higgs, bottom, and top
quark supermultiplets.  Note that the RGE's now become dependent on
all of the different gaugino couplings.  (See, for example, Appendix B
of Ref.~\cite{chankowski}.)  However, such corrections from
differences in the couplings are higher order effects, and may be
neglected here.

An interesting case in which Yukawa couplings could be important is in
theories with $R_P$ violation.  In the minimal supersymmetric standard
model, lepton and baryon number are not accidental symmetries of the
low-energy theory, but are put in by hand when one imposes $R_{P}$
conservation.  $R_P$-violating terms include Yukawa couplings of
leptons $\lambda LLE$, lepton doublets to quarks $\lambda'LQD$, and
the different quark singlets $\lambda''UDD$, where generational
indices have been suppressed. Current bounds on individual couplings
allow rather large couplings $\lambda$, $\lambda'$, and $\lambda''$
for certain generational indices.  (See, for example,
Ref.~\cite{erler}, where present constraints on $\lambda'_{333}$, the
only coupling with three third generation indices, are analyzed.)  In
addition, these bounds are often significantly weakened for heavy
superpartner masses, and so, in the scenarios we are considering, may
be extremely poor. Consequently, important negative and
flavor-dependent Yukawa contributions to $\susyU_i^f$ could arise in
$R_{P}$-violating models.  Of course, $R_P$ violation also allows the
lightest supersymmetric particle to decay, leading to non-standard
supersymmetric signals, which modifies the strategies for measuring
such super-oblique parameters.

In the absence of flavor conservation, flavor mixing matrix elements
will appear at the gaugino-fermion-sfermion vertices.  In this case,
if a sfermion in one generation belongs to the heavy sector and a
sfermion in another generation belongs to the light sector, as may be
the case, for example, in 2--1 models, heavy sfermion loops may appear
in the matter wavefunction and vertex renormalizations of the gaugino
couplings of the light sector through flavor-violating interactions.
Such effects also contribute to the violation of the Ward identity for
gaugino couplings. However, in such models, flavor mixings between the
heavy and the light sectors are naturally suppressed by
$m/M$. Therefore, the effects of these flavor-violating loop
corrections should be small.  Note, however, that such mixings may be
measured or bounded by experiment\cite{ACFH}, and such effects have
implications for gaugino coupling measurements\cite{gex}.

Up to this point, we have only discussed deviations of the SUSY
relation between the gauge couplings and the gaugino couplings. In
supersymmetric theories, there are also $D$-term quartic scalar
couplings, which arise from SUSY gauge interactions, and are therefore
proportional to $g_i^2$ in the SUSY limit.  After the heavy
superpartners decouple, the relations between the quartic scalar
couplings and the gauge couplings also receive non-decoupling
corrections (which can be viewed as super-oblique corrections from the
wavefunction renormalization of the auxiliary $D$ fields), and also
possibly the flavor-dependent corrections discussed above. However,
such deviations are likely to be more difficult to investigate
experimentally: the couplings of four physical scalars are extremely
challenging to measure, and other probes of $D$-terms, such as in
Higgs decays and SU(2) doublet sfermion splitting, require ambitious
measurements of other parameters, such as $\tan\beta$, the ratio of
Higgs expectation values.  Although such measurements may be possible
in certain scenarios, in the rest of this study, we will concentrate
on the super-oblique corrections between the gauge couplings and
gaugino couplings, which enter generically in all processes involving
gauginos, and which appear much more promising experimentally.

\section{Numerical Estimates in Theoretical Frameworks}
\label{sec:framework}

In Sec.~\ref{sec:decoupling}, we discussed super-oblique corrections
in the general context of models with heavy and light sectors with
arbitrary particle content.  In this section, we will investigate what
size corrections may be reasonably expected.  For concreteness, we
will consider the two well-motivated classes of models described in
Sec.~\ref{sec:introduction}, namely, ``heavy QCD models,'' in which
the heavy sector includes all colored superpartners, and ``2--1
models,'' in which the heavy sector consists of the first two families
of sfermions.  We will estimate the contributions of the heavy sectors
to the parameters $\susyU_{i}$ and $\dsusyU_{ij}$ in these two
frameworks, treating all heavy sector particles as degenerate ---
non-degeneracies within the heavy sector typically only lead to higher
order effects.  Discussion of additional contributions to $\susyU_i$
in models that contain vector-like multiplets at some high mass scale,
{\em e.g.}, in gauge mediation and U(1)$'$ models, is deferred to
Sec.~\ref{sec:messenger}.  Note that while the results of this section
are presented to serve as benchmarks, it is important to keep in mind
that much larger effects may be possible from, for example, exotic
particles.

\subsection{Heavy QCD models}
\label{sec:heavyQCD}

We first consider models with all strongly-interacting sparticles in
the heavy sector.  This category includes models in which the sfermion
and gaugino masses are dominated by a flavor-independent term that is
a function of the low-energy gauge couplings.  The hierarchy between
the strong and electroweak gauge couplings is then translated into a
mass hierarchy between colored and non-colored particles. Examples
include the no-scale limit of minimal supergravity\cite{noscale}, in
which scalar masses are determined only by gaugino loops, models with
non-universal gaugino masses and a heavy gluino\cite{ACM}, in which
squark masses are enhanced by gluino loop contributions, and
gauge mediation models\cite{gm}, in which the gaugino and sfermion
masses are determined by gauge loops involving vector-like messenger
supermultiplets at the $\sim {\cal O}(100 \text{ TeV})$ scale.

In these models, minimization of the Higgs potential implies, given
the constraint of the $Z$ boson mass, that the Higgsino mass parameter
$\mu$ is naturally of the order of the gluino mass.  Thus, typically
the Higgsinos and one Higgs doublet should be included in the heavy
sector.  However, the contributions of these particles to $\susyU_{i}$
and $\dsusyU_{ij}$ are small, and the primary impact of the scale of
$\mu$ is on what experimental observables may be available to probe
the super-oblique corrections.

Assuming that the heavy sector consists of all squarks and the gluino,
we present in Table~\ref{table:t1} the $\beta$-function coefficients
and the resulting parameters $\susyU_i$ from
Eq.~(\ref{delta}).\footnote{The contribution of a (component) field
$j$ with spin $S^j$ to the $\beta$-function coefficient $b_{i}$ is
$b_i^j = N_i^j a^j T_i^j$, where $N_i^j$ is the appropriate
multiplicity; $a^j = \frac{1}{3}, \frac{2}{3}, -\frac{11}{3}$ for $S^j
= 0, \frac{1}{2}, 1$, respectively; and $T_i^j = 0, \frac{1}{2}, 2,3$,
or $\frac{3}{5}Y^2$ for a singlet, a particle in the fundamental
representation of $SU(N)$, an $SU(2)$ triplet, an $SU(3)$ octet, or,
for $i = 1$, a particle with hypercharge $Y=I-Q$, respectively.}
Inclusion of the Higgsinos and one Higgs doublet in the heavy sector
would slightly enhance $\susyU_1$ and $\susyU_2$.  We have chosen the
grand-unification normalization for the hypercharge U(1); of course,
$\susyU_1$ is independent of this choice.  For simplicity, we assume
$g_i(m) = g_i(M_Z)$ in our numerical estimates, which is sufficient
for $m/M_Z \alt 3$. We also have

\begin{equation}
\dsusyU_{21} \approx \frac{1}{16\pi^2}
\left[3g_2^2(m) - \frac{11}{5} g_1^2(m)\right]\ln R \approx 
0.50\% \times \ln R \ .
\label{rho21a}
\end{equation}
The parameters $\susyU_i$ and $\dsusyU_{21}$ are logarithmically
dependent on $R=M/m$; a typical value for this ratio in heavy QCD
models is $R\sim {\cal O}(10)$.

\begin{table}
\caption{The $\beta$-function coefficients and parameters $\susyU_{i}$ 
in the heavy QCD models. $R\equiv M/m$ is the ratio of heavy to light
mass scales.}
\begin{tabular}{ccccc}
Gauge Group $G_{i}$ & $b_{g_i}$  & $b_{h_i}$ & $b_{g_i} - b_{h_i}$
& $\susyU_i$ \\
\hline 
SU(3) & $-7$ & frozen & $-7$& $-6.7\% \times \ln R$ \\
SU(2) & $-\frac{1}{2}$ & $-\frac{7}{2}$ & $3$ & 
$0.80\% \times \ln R$ \\
$\frac{5}{3}$U(1) & $\frac{11}{2}$ & $\frac{33}{10}$ &  
$\frac{11}{5}$ &  $0.29\% \times \ln R$ \\
\end{tabular}
\label{table:t1}
\end{table}

In these models, the gluino and the squarks are in the heavy sector
and are decoupled.  The coupling $h_3$ is therefore not renormalized
below $M$, and by convention, we take its value below $M$ to be
frozen, with $h_3(Q<M) = h_3(M) = g_3(M)$.  By assumption, the gluino
and squarks are inaccessible at colliders, and so the parameter
$\susyU_3$ may be measured only through their virtual effects. Such
measurements are likely to be extremely difficult, as they require an
understanding of process-dependent QCD corrections. Note, however,
that if the gluino is light, then $b_{g_3} - b_{h_3} = 4$.  Hence, the
sign of $\susyU_{3}$ could offer an indirect test of the
${\cal{O}}$(GeV) gluino scenario \cite{farrar} if both gluinos and
squarks are not observed at the LHC.

In the expressions above, we have treated all gaugino couplings as
equivalent.  In fact, as discussed in Sec.~\ref{sec:other}, the
various gaugino couplings may be significantly differentiated by
Yukawa couplings.  In this case, the gaugino-Higgsino-Higgs couplings
$h_i^H$ are split from the other gaugino couplings by non-decoupling
corrections from the heavy $t$ and $b$ squarks.  The corresponding
parameters $\susyU_i^H$ are therefore diminished by the effects of the
$t$- and $b$-quark Yukawa couplings and may be large and negative.
For the remaining couplings, we have explicitly confirmed by
comparison with the complete set of one-loop RGE's for heavy QCD
models contained in Ref.~\cite{chankowski} that the additional
decoupling effects not included in Eq.~(\ref{delta}) are negligible.

\subsection{2--1 models}

Models with heavy first two generation scalars and light third
generation scalars, Higgs fields, and gauginos have been discussed in
Ref.~\cite{effectivesusy}, with explicit examples given recently in
Refs.~\cite{u1,cohen,nilles}.  These models exploit and are motivated
by the fact that the most stringent flavor-violating constraints may
be satisfied by taking the sfermions of the first two families very
heavy, while fine-tuning concerns may be alleviated by taking the
other sparticles light, since the Higgs sector couples (at leading
order) only to the sfermions of the third family and the electroweak
gauginos.  Note, however, that the heavy scale propagates to the light
fields via hypercharge $D$-terms and two-loop effects, leading to a
strongly model-dependent upper limit on the heavy scale
$M$\cite{effectivesusy,badfine}.  Typical values of $R \sim 40 - 200$
may be taken in these models.

The gluino could, in principle, belong to either sector.  For
definiteness and motivated by gaugino mass unification, we will assume
that all gauginos are in the light sector.  The resulting parameters
$\susyU_i$ are given in Table~\ref{table:t2}.  Since the decoupled
sector consists of complete multiplets of a grand unified group, the
differences $b_{g_i} - b_{h_i}$ are equal for all $i$, and the
expressions for $\dsusyU_{ij}$ are simplified:

\begin{eqnarray}
\label{rho32}
\dsusyU_{32} &\approx& \frac{1}{16\pi^2} 
\frac{8}{3}\left[g_3^2(m) - g_2^2(m)\right]
\ln R \approx 1.8\% \times \ln R \ , \\
\label{rho31}
\dsusyU_{31} &\approx& \frac{1}{16\pi^2}
\frac{8}{3}\left[g_3^2(m) - g_1^2(m)\right]
\ln R \approx 2.2\% \times \ln R \ , \\
\label{rho21b}
\dsusyU_{21} &\approx& \frac{1}{16\pi^2}
\frac{8}{3}\left[g_2^2(m) - g_1^2(m)\right]
\ln R \approx 0.35\% \times \ln R \ .
\end{eqnarray}
We see that the parameters $\susyU_3$, $\dsusyU_{32}$, and
$\dsusyU_{31}$ are enhanced by the strong coupling and are therefore
promising observables to probe.

\begin{table}
\caption{The parameter $\susyU_{i}$ in the 2--1 models.}
\begin{tabular}{ccccc}
Gauge Group $G_{i}$ & $b_{g_{i}}$  & $ b_{h_{i}}$ & 
$b_{g_{i}} - b_{h_{i}}$ & $\susyU_{i}$ \\
\hline 
SU(3) & $-\frac{13}{3}$ & $-7$ & $\frac{8}{3}$& $2.5\% \times \ln R$ \\
SU(2) & $-\frac{1}{3}$ & $-3$ & $\frac{8}{3}$& $0.71\% \times \ln R$ \\
$\frac{5}{3}$U(1) & $\frac{158}{30}$ & $\frac{78}{30}$ & 
$\frac{8}{3}$ &  $0.35\% \times \ln R$ \\
\end{tabular}
\label{table:t2}
\end{table}

Variants of 2--1 models may give alternative mass patterns, such as,
for example, light and degenerate left-handed sleptons of the first
two generations, and heavy right-handed selectrons and
smuons\cite{nelson}.  A generalization of our results to these cases
is straightforward.  A reduced heavy sector diminishes $b_{g_{i}} -
b_{h_{i}}$ and, thus, the corrections $\susyU_{i}$ and $\dsusyU_{ij}$.
On the other hand, the existence of light selectrons and electron
sneutrinos more than makes up for this setback, as it opens up the
possibility of high precision probes of the electroweak super-oblique
parameters at $e^{\pm} e^-$ colliders that are inaccessible if these
sleptons are all heavy\cite{slepton,gex}.

\section{Vector-like (messenger) sectors}
\label{sec:messenger}

The super-oblique parameters receive contributions from all split
supermultiplets with standard model quantum numbers.  In many SUSY
extensions of the standard model, there are extra vector-like fields
which transform under the standard model gauge groups. These
vector-like fields could have both SUSY preserving and SUSY breaking
masses, and so they can also contribute to deviations in the SUSY
relations $g_i=h_i$ at low energies.  For example, this is the case in
the gauge-mediated SUSY breaking models, where the vector-like
messenger sector contains Dirac fermions with mass $M_V$ and complex
scalars with squared masses $M_V^2(1\pm x)$. The low-energy ordinary
sfermion spectrum is determined by $M_V$ and $x$, and it is required
that $|x|<1$ in order to avoid tachyons and contradiction with
experiments. More generally, irrespective of the mechanism that
mediates SUSY breaking to the ordinary sector, there could exist some
exotic vector-like fields at or above the weak scale with Dirac
fermions with mass $M_V$ and complex scalars with squared masses
$M_V^2(1+x)$ and $M_V^2(1+y)$.  The variables $x$ and $y$ represent
the SUSY breaking effects.  If SUSY breaking is mediated through
supergravity, $x$ and $y$ can be ${\cal O}(1)$ only when the
vector-like fields have masses near the weak scale.  If SUSY breaking
is mediated through gauge interactions, $x$ and $y$ may be ${\cal
O}(1)$ only when the vector-like fields are $\alt {\cal O} (100 \text{
TeV})$; otherwise, through loop corrections, they will generate SUSY
breaking masses for standard model superpartners that are too large.

We consider first the messenger fields of gauge-mediated SUSY breaking
models. Let $b_i$ be the contribution of the entire vector-like
supermultiplet sector to the appropriate one-loop $\beta$-function
coefficient.  For example, if the messenger sector contains $n_5$
pairs of $5$ and $\overline{5}$ and $n_{10}$ pairs of $10$ and
$\overline{10}$ SU(5) multiplets, then $b_{i} = n_{5} + 3n_{10}$ for
all $i$.  If we naively perform a leading logarithm calculation,
thereby ignoring finite pieces and decoupling all loops at the mass of
the heaviest particle in the loop, we find

\begin{equation}
\delta \susyU_i^{\text{LL}} \approx 
\frac{g_i^2(M_V)}{64\pi^{2}}b_i \left(\frac{2}{3} \ln \sqrt{1+x} -
\frac{1}{3}\ln \sqrt{1-x} \right) \ .
\end{equation}
As is evident from this expression, the leading logarithms $\ln
(M_V/\mu)$ have cancelled, as they must, since in this case, the SUSY
breaking is governed not by $M_V$, but by $x$.  The result is
therefore reduced to a finite term, and we clearly must calculate the
finite pieces correctly.

For gauge couplings, the naive decoupling is correct: the scalar loops
decouple at $M_V \sqrt{1\pm x}$ and the fermion loops decouple at
$M_V$. (See the Appendix.) The contribution to the gauge couplings can
be written as

\begin{equation}
\delta \left(\frac{1}{g_i^2} \right) 
= - \frac{b_i}{8\pi^2} \frac{1}{2}\left[ \frac{1}{3}
\ln \frac{M_V \sqrt{1+x}}{\mu} +2\times \frac{2}{3} \ln \frac{M_V}{\mu}
+ \frac{1}{3} \ln \frac{M_V\sqrt{1-x}}{\mu} \right] \ .
\label{gevolution}
\end{equation}
For the fermion-sfermion loop contribution to the gaugino wavefunction
renormalization, we can apply the result in the Appendix. The
Feynman parametrization integral of Eq.~(\ref{Feynmanintegral})
becomes

\begin{eqnarray}
\lefteqn{\int_0^1 d\alpha 2\alpha \ln \left[\frac{\alpha M_V^2(1\pm x)+
(1-\alpha)M_V^2}{\mu^2} \right]} \nonumber \\ 
&&\quad = \ln \frac{M_V^2}{\mu^2} - \frac{1}{2} \pm \frac{1}{x} + 
\ln (1\pm x) -\frac{1}{x^2} \ln (1\pm x)\ ,
\end{eqnarray}
and so the contribution to the gaugino couplings is

\begin{eqnarray}
\delta \left(\frac{1}{h_i^2} \right) 
= - \frac{b_i}{8\pi^2} \frac{1}{2}\left[\left(
\ln \frac{M_V}{\mu} -\frac{1}{4} +\frac{1}{2x} +\frac{1}{2}
\ln(1+x) -\frac{1}{2x^2} \ln(1+x) \right) \right. \nonumber \\
\left. +\left( \ln \frac{M_V}{\mu}
-\frac{1}{4} -\frac{1}{2x} +\frac{1}{2}
\ln(1-x) -\frac{1}{2x^2} \ln(1-x) \right) \right] \ .
\label{hevolution}
\end{eqnarray}
As expected, we find that the $\ln ({M_V}/{\mu})$ terms cancel in the
difference between the $g_i$ and $h_i$ evolutions given in
Eqs.~(\ref{gevolution}) and (\ref{hevolution}), and the final result
is

\begin{eqnarray}
\delta \susyU_{i} &=& \frac{g_{i}^{2}(M_V)}{16\pi^{2}}b_{i}\left[
-\frac{1}{4} + \left(\frac{1}{6} - \frac{1}{4x^{2}}\right)\ln\left(
1 - x^{2}\right)\right] \nonumber \\
&\approx& -\frac{g_i^2(M_V)}{384\pi^{2}}b_{i}x^{2} \ , \quad
\text{for small $|x|$} \ .
\label{messenger}
\end{eqnarray}
The effect is very small for most of the range $0\le |x|<1$, and it is
therefore unlikely that any experimental measurement can be sensitive
to super-oblique corrections arising from such a messenger sector.
Note, however, that this effect has a negative sign for small $x$
relative to the logarithmic effect discussed in
Secs.~\ref{sec:decoupling} and \ref{sec:framework}.  Its smallness is
thus fortunate, in the sense that such effects therefore cannot cancel
the non-decoupling signatures of heavy superpartners.
 
It is also straightforward to obtain the result for the more general
spectrum of vector-like fields ($x \neq -y$):

\begin{eqnarray}
\delta \susyU_{i} &=& \frac{g_{i}^{2}(M_V)}{16\pi^{2}}b_{i}\left[
-\frac{1}{8} + \frac{1}{4x}+\left( \frac{1}{6} 
-\frac{1}{4x^2} \right) \ln(1+x) \right] + \left( x\to y \right)
\nonumber \\
&\approx& \frac{g_{i}^{2}(M_V)}{16\pi^{2}}b_{i}\left(
\frac{x+y}{12} - \frac{x^2+y^2}{48} \right)\ , \quad
\text{for small $|x|$ and $|y|$} \ .
\end{eqnarray}
For $x\neq -y$, the linear term does not vanish and we have a larger
effect. However, unless there are many such heavy vector-like
multiplets (large $b_i$) with significant mass splittings among
supermultiplet components (large $|x|$, $|y|$), the contributions to
the super-oblique corrections are small relative to the deviations
discussed in Sec.~\ref{sec:framework}. Note that in both the case of
vector-like messenger sectors and this more general case, large
deviations are possible only for $|x|, |y| \approx 1$.  If a deviation
is seen which cannot be due simply to the standard model
superpartners, the considerations stated above then strongly suggest
that the masses of such vector-like particles are below the ${\cal
O}(100 \text{ TeV})$ scale.

\section{Final Comments and Conclusions}
\label{sec:conclusions}

In this study we have considered low-energy softly broken
supersymmetric theories that contain a heavy sparticle sector that is
beyond the kinematical reach of planned future collider experiments.
Sparticle spectra leading to such scenarios appear in certain limits
of the most simple supergravity model, but more importantly, are known
to arise in many other well-motivated frameworks for the soft SUSY
breaking parameters, and especially those that address the SUSY flavor
problem.  Here, we have shown that the heavy sparticle sector induces
non-decoupling radiative corrections in the light sparticle sector,
providing a crucial window for the exploration of the heavy sector
through precision measurements in processes involving light
sparticles.

The non-decoupling of SUSY breaking is analogous to the non-decoupling
of SU(2) breaking in the standard model.  Here we have considered a
particularly important set of non-decoupling effects, which are
analogous to the oblique corrections of the standard model, and which
we therefore call super-oblique corrections.  Such corrections arise
from gauge boson and gaugino wavefunction renormalization, and lead to
deviations in the equivalence of gauge boson couplings $g_i$ and
gaugino couplings $h_i$.  These corrections are therefore most closely
identified with the oblique parameter $U$, and we have parametrized
them with the super-oblique parameters $\susyU_i \equiv h_i/g_i-1$.
The super-oblique parameters have a number of important features: they
are model-independent measures of SUSY breaking, receive additive
contributions from every split supermultiplet, and grow
logarithmically with $M/m$, the ratio of heavy to light mass scales.

The super-oblique parameters may be expressed simply in terms of
$\ln(M/m)$ and group theory factors.  As examples, we have estimated
the corrections from heavy superpartners within specific theoretical
frameworks and found typical values $\susyU_i \approx P_i \ln(M/m)$,
where $P_i = 0.3\%, 0.7\%, 2.5\%$ for $i=1,2,3$, and the logarithm
varies between 2 and 5. The hierarchy between the different parameters
results from their proportionality to the low-energy gauge couplings,
and the positive sign of the parameters is model independent at the
leading logarithm level.  We also calculated the contributions of
messenger sectors in models of gauge mediation and possible exotic
vector-like multiplets.  Such contributions were found to be typically
very small, with substantial corrections only for highly split
multiplets.

The effect of super-oblique corrections in the accessible sparticles
is to modify gaugino coupling constants.  It is therefore not
difficult to identify observables that are formally probes of such
corrections.  For example, the cross section of chargino production at
$\epem$ colliders provides one such observable\cite{FMPT}, as the
gaugino couplings $h_2$ enter through $t$-channel sneutrino exchange.
Selectron production at an $\epem$ collider provides another such
probe\cite{slepton}.  In addition, if a particle has two or more decay
modes, and at least one involves gauginos, its branching ratios are
also probes of the super-oblique corrections.  Of course, all such
measurements receive uncertainties from a variety of sources, ranging
from backgrounds and finite statistics to the errors arising from the
many other unknown SUSY parameters entering any given process.  A
classification of possible experimental probes at $e^{\pm} e^-$ and
hadron colliders, as well as detailed studies of promising
measurements incorporating such uncertainties, is contained in an
accompanying article\cite{gex}.

If super-oblique corrections are measured, the implications are many
and varied, depending on what precision is achieved and what scenario
is realized in nature.  The implications may be listed in increasing
order of the precision of the measurements.  If super-oblique
parameters are constrained to be roughly consistent with zero, such
tests provide quantitative confirmation that such particles are indeed
supersymmetric particles.  If bounds on $\susyU_i$ at the level of
$P_i\%\times \ln (M/m)$ are achieved, deviations from zero may be
seen, providing evidence of a heavy sector.  Finally, if bounds at the
level of $P_i\%$ are achieved, the heavy mass scale may be constrained
to within a factor of 3, providing a discriminant for model building,
and in the most optimal scenarios, setting a target for future
collider searches.  Alternatively, if all superpartners are
directly observed, deviations from $g_i = h_i$ are indications of the
existence of, for example, exotic matter with highly split
supermultiplets, which are likely to be below the ${\cal O} (100
\text{ TeV})$ scale.  If supersymmetry is discovered, the
super-oblique corrections will therefore provide a crucial window on
the physics above the TeV scale.

\acknowledgements

The authors are grateful to L.~Hall, M.~Nojiri, M.~Peskin, and
J.~Terning for conversations.  This work was supported in part by the
Director, Office of Energy Research, Office of High Energy and Nuclear
Physics, Division of High Energy Physics of the U.S.  Department of
Energy under Contracts DE--AC03--76SF00098 and DE--AC02--76CH03000,
and in part by the NSF under grants PHY--95--14797 and PHY--94--23002.
J.L.F. is supported by a Miller Institute Research Fellowship and
thanks the high energy theory group at Rutgers University for its
hospitality.

While completing this work, we learned of related work in
progress\cite{otherwork}.  We thank D.~Pierce, L.~Randall, and
S.~Thomas for bringing this work to our attention.

\appendix
\section*{One-loop threshold corrections at the heavy superpartner
mass scale}
\label{sec:a1}

In this appendix, we calculate the one-loop threshold corrections at
the heavy superpartner scale. These finite corrections are usually
included only when one uses 2-loop RGE's. However, since in our case
$\ln (M/m)$ is not necessarily very large, it is not clear {\em a
priori} that the finite pieces are negligible relative to the leading
logarithm contributions. It is therefore important that we consider
these pieces in detail.  This will be seen to be especially true when
we consider the contributions from vector-like messenger fields in
models of gauge-mediated SUSY breaking, where the large logarithms
cancel and the finite pieces must be treated carefully. This is
discussed in Sec.~\ref{sec:messenger}.

In calculating these corrections, we work in the SUSY preserving
$\overline{\text{DR}}$ renormalization scheme, since we want to
preserve the relation $g_i=h_i$ when SUSY is not broken.\footnote{In
fact, our calculation is the same as in the $\overline{\text{MS}}$
scheme, as we only have scalars and fermions in the loop.} The
couplings measured at low energies should be converted into the same
scheme before comparison.

We first consider the vacuum polarization of the gauge bosons due
to the heavy scalar loops, $\Pi_H^{\mu \nu}(q)= (g^{\mu \nu}q^2
-q^{\mu} q^{\nu}) \Pi_H(q^2)$. The couplings are measured at 
much lower energies than the heavy scalar mass $M_S$, so we set
the external momentum $q$ to zero. The vacuum polarization is then
given by the well-known result

\begin{eqnarray}
\Pi_H(0) &=& i g^2 \mu^{4-d} T_R \int_{0}^{1} d\alpha
 \int \frac{d^d k}{(2\pi)^d} \frac{(1-2\alpha)^2}{(k^2
 - M_S^2)^2}  \nonumber \\
&=& -\frac{T_R}{3} \frac{g^2}{16\pi^2} \left( \frac{1}{2-
 \frac{d}{2}} - \gamma_E + \ln 4\pi - \ln \frac{M_S^2}{\mu^2} \right)
 + {\cal O}(4-d) \ , 
\end{eqnarray}
where here $\mu$ is the renormalization scale. $T_R$ is defined by
$T_R \delta^{ab} = \text{tr}\; T^a T^b$ and is $\frac{1}{2}$ for the
fundamental representation of SU(N).  We subtract the terms
$1/(2-\frac{d}{2}) - \gamma_E+\ln 4\pi$ in the $\overline{\text{DR}}$
scheme. The remaining term, $\ln (M_S^2/ \mu^2)$, vanishes when
$\mu=M_S$, implying that the gauge coupling in the low-energy
effective theory matches that in the high energy theory at $\mu=M_S$.
Therefore, we decouple the heavy scalar loops at the scale of their
masses in calculating the low-energy gauge boson couplings. In doing
so, there is no finite threshold correction at one-loop.

Now we turn to the low-energy gaugino couplings. The heavy loop of the
gaugino wavefunction renormalization consists of a scalar and a
fermion of masses $M_S$ and $m_f$, respectively. The one-loop diagram
gives

\begin{eqnarray}
\Sigma_{2H}(q) &=& i (-i\sqrt{2} h)^2 \mu^{4-d}
 T_R \int \frac{d^d k}{(2\pi)^d}
 \frac{i(\not{k} +m_f)}{k^2-m_f^2} \frac{i}{(k-q)^2- M_S^2}
\nonumber \\
&=& i 2 h^2 \mu^{4-d}
 T_R \int_0^1 d\alpha \int \frac{d^d k}{(2\pi)^d}
 \frac{\alpha\not{q} +m_f}{[k^2 +\alpha(1-\alpha)q^2 -\alpha M_S^2
 -(1-\alpha)m_f^2]^2} \ .
\end{eqnarray}
Setting the external momentum to zero, the contribution to the
wavefunction renormalization is

\begin{eqnarray}
\delta Z_2 &=& \left. \frac{d \Sigma_{2H}}{d \not{q}}
 \right|_{\not{q} \to 0} = i 2 h^2 \mu^{4-d}
 T_R \int_0^1 d\alpha
 \int \frac{d^d k}{(2\pi)^d} \frac{\alpha}{[k^2-\alpha M_S^2
 -(1-\alpha)m_f^2]^2} \nonumber \\
\label{Feynmanintegral}
&=& - T_R \frac{h^2}{16\pi^2} \int_0^1 d\alpha 2\alpha \left[
 \frac{1}{2-\frac{d}{2}} -\gamma_E +\ln 4\pi 
 -\ln \frac{\alpha M_S^2 +(1-\alpha) m_f^2}{\mu^2} \right]
 + {\cal O}(4-d) \ .
\end{eqnarray}
For the fermion-sfermion loop, $m_f\simeq 0$, and the Feynman integral
reduces to

\begin{equation}
\int_0^1 d \alpha 2\alpha \ln \frac{\alpha M_S^2}{\mu^2}
=\ln \frac{M_S^2}{\mu^2} -\frac{1}{2} = \ln \left(
\frac {M_S e^{-\frac{1}{4}}}{\mu} \right)^2 \ .
\end{equation}
In this case, there is a nonzero finite correction, which implies that
the decoupling scale of the fermion-sfermion loop is at $M_S
e^{-\frac{1}{4}}$ instead of $M_S$.\footnote{In the heavy
Higgsino-light Higgs case, we have $\int_0^1 d \alpha 2\alpha \ln
\frac{(1-\alpha) m_{\tilde{H}}^2}{\mu^2} =\ln \frac{m_{\tilde{H}}^2}
{\mu^2} -\frac{3}{2}$.}  Therefore, to take account of the threshold
corrections at the decoupling scale, we could replace the scale $M$ in
Eq.~(\ref{h}) by an effective decoupling scale $\widetilde{M}$ 
different from that in Eq.~(\ref{g}).

To get an understanding of how large such shifts in the decoupling
scale are, let us consider theories with heavy sectors composed of
scalars (and possibly gauginos) with mass $M$.  Including the one-loop
threshold corrections, we have

\begin{equation}
\frac{1}{h_{i}^{2}(m)}  \approx
\frac{1}{h_{i}^{2}(M)} + \frac{b_{i}}{8\pi^2} \frac{1}{4}
+ \frac{b_{h_{i}}}{8\pi^{2}}\ln \frac{M}{m}
- \frac{b_{h_i}}{8\pi^2} \frac{1}{4} \ ,
\end{equation}
where $b_{i}$ is the one-loop $\beta$-function coefficient for both
gauge and gaugino couplings above the squark scale. The deviation of
$h_i$ from $g_i$ at low energies becomes

\begin{eqnarray}
\frac{h_{i}(m)}{g_{i}(m)} & \approx & 1 
+ \frac{g_{i}^{2}(m)}{16\pi^{2}}(b_{g_{i}} - b_{h_{i}})
\ln\frac{M}{m} -\frac{g_{i}^{2}(m)}{16\pi^{2}}(b_{i} - b_{h_{i}}) 
\frac{1}{4}  \nonumber \\
&=& 1+ \frac{g_{i}^{2}(m)}{16\pi^{2}}(b_{g_{i}} - b_{h_{i}})
\left( \ln \frac{M}{m} -\frac{3}{8} \right)\ .
\end{eqnarray}
Here we have used the relation $b_{i}-b_{h_i} = \frac{3}{2}
(b_{g_i}-b_{h_i})$, valid since $b_{i}-b_{h_i}$ receives contributions
from heavy scalars and their fermionic partners, while
$b_{g_i}-b_{h_i}$ receives contributions only from the fermionic
partners. We can see that the deviation is slightly smaller than that
naively obtained by decoupling the heavy loop at the heaviest particle
mass. However, as we are interested in the case where $\ln (M/m)
\agt 2$, the shift only introduces only a small correction to 
the total deviation.



\begin{thebibliography}{99}

\bibitem{finetuning}
J.~Ellis, K.~Enqvist, D.~V.~Nanopoulos, and F.~Zwirner,
Mod.~Phys.~Lett.~A {\bf 1}, 57 (1986); R.~Barbieri and G.~F.~Giudice,
Nucl.~Phys.~B {\bf 306}, 63 (1988); G.~W.~Anderson and D.~J.~Castano,
Phys.~Lett.~B, {\bf 347}, 300 (1995); {\em ibid.}, Phys.~Rev.~D {\bf
52}, 1693 (1995); {\em ibid.}, Phys.~Rev.~D {\bf 53}, 2403 (1996).

\bibitem{LHC}
CMS Collaboration, Technical Proposal, CERN/LHCC/94--38 (1994);
ATLAS Collaboration, Technical Proposal, CERN/LHCC/94--43 (1994).

\bibitem{JLC} 
JLC Group, {\it JLC--I}, KEK Report No. 92--16, Tsukuba (1992).

\bibitem{NLC}
NLC ZDR Design Group and the NLC Physics Working Group, S.~Kuhlman
{\em et al.}, Physics and Technology of the Next Linear Collider,
hep-ex/9605011;
The NLC Design Group, C.~Adolphsen {\em et al.}, Zeroth-Order Design
Report for the Next Linear Collider, LBNL--PUB--5424, SLAC Report No.
474, UCRL--ID--124161 (1996).

\bibitem{DESYLC}
ECFA/DESY LC Physics Working Group, E.~Accomando {\em et al.},
DESY--97--100, hep-ph/9705442.

\bibitem{gm}
M.~Dine, W.~Fischler, and M.~Srednicki, Nucl.~Phys. {\bf B189}, 575
(1981);
C.~Nappi and B.~Ovrut, Phys.~Lett.~B {\bf 113}, 175 (1982);
M.~Dine and W.~Fischler, Nucl.~Phys. {\bf B204}, 346 (1982);
L.~Alvarez-Gaume, M.~Claudson and M.~Wise, Nucl.~Phys. {\bf B207}
96 (1982); 
M.~Dine, A.~Nelson, and Y.~Shirman, Phys.~Rev.~D {\bf 51}, 
1362 (1995);
M.~Dine, A.~Nelson, Y.~Nir, and Y.~Shirman, Phys.~Rev.~D {\bf 53}, 
2658 (1996).

\bibitem{noscale}
A.~B.~Lahanas and D.~V.~Nanopoulos, Phys.~Rept. {\bf 145}, 1 (1987). 

\bibitem{ACM}
N.~Arkani-Hamed, H.--C.~Cheng, and T.~Moroi, Phys.~Lett.~B {\bf 387},
529 (1996).

\bibitem{effectivesusy}
S.~Dimopoulos and G.~F.~Giudice, Phys.~Lett.~B {\bf 357}, 573 (1995); 
A.~Pomarol and D.~Tommasini, Nucl.~Phys.~{\bf B466}, 3 (1996).

\bibitem{u1}
G.~Dvali and A.~Pomarol, Phys.~Rev.~Lett.~{\bf 77}, 3728 (1996); 
R.~N.~Mohapatra and A.~Riotto, Phys.~Rev.~D {\bf 55}, 1 (1997);
R.-J.~Zhang, JHU--TIPAC--97004, hep-ph/9702333.

\bibitem{cohen}
A.~G.~Cohen, D.~B.~Kaplan, and A.~E.~Nelson, Phys.~Lett.~B {\bf 388},
588 (1996);
A.~G.~Cohen, D.~B.~Kaplan, F.~Lepeintre, and A.~E.~Nelson,
Phys.~Rev.~Lett. {\bf 78}, 2300 (1997);
A.~E.~Nelson and D.~Wright, UW/PT--97--03, hep-ph/9702359.

\bibitem{nilles}
H. P. Nilles and N. Polonsky, RU--97--47;
N. Polonsky, talk presented at Supersymmetry 97, 
Fifth International Conference on Supersymmetries in Physics,
The University of Pennsylvania, Philadelphia, May 27--31, 1997.

\bibitem{badfine}
N.~Arkani-Hamed and H.~Murayama, hep-ph/9703259.

\bibitem{finetuning2}
P.~Ciafaloni and A.~Strumia, FT--UAM--96--43, hep-ph/9611204;
K.~Agashe and M.~Graesser, LBNL--40121, hep-ph/9704206.

\bibitem{Peskin} 
M.~Peskin and T.~Takeuchi, Phys. Rev. Lett. {\bf 65}, 964 (1990);
{\em ibid.}, Phys.~Rev.~D {\bf 46}, 381 (1992).

\bibitem{oblique}
B.~Holdom and J.~Terning, Phys.~Lett.~B {\bf 247}, 88 (1990);
M.~Golden and L.~Randall, Nucl.~Phys.~{\bf B361}, 3 (1991); 
G.~Altarelli and R.~Barbieri, Phys.~Lett.~B {\bf 253}, 161 (1991);
G.~Altarelli, R.~Barbieri, and S.~Jadach, Nucl.~Phys.~{\bf B269},
3 (1992).

\bibitem{FMPT}
J.~L.~Feng, H.~Murayama, M.~E.~Peskin, and X.~Tata, Phys.~Rev.~D {\bf
52}, 1418 (1995).

\bibitem{slepton}
M.~M.~Nojiri, K.~Fujii, and T.~Tsukamoto, Phys.~Rev.~D {\bf 54}, 6756
(1996).

\bibitem{Hikasa}
K.~Hikasa and Y.~Nakamura, Z.~Phys.~C {\bf 70}, 139 (1996); {\em
ibid.}, {\bf 71}, 356 (1996).

\bibitem{gex}
H.-C.~Cheng, J.~L.~Feng, and N.~Polonsky, FERMILAB--PUB--97/205--T,
LBNL--40466, UCB--PTH--97/34, RU--97--46.

\bibitem{AC} 
T.~Appelquist and J.~Carrazone, Phys.~Rev.~D {\bf 11}, 2856 (1975).

\bibitem{threshold1}
See, for example, P.~Langacker and N.~Polonsky, Phys.~Rev.~D {\bf 47},
4028 (1993); 
{\em ibid.}, {\bf 52}, 3081 (1995).

\bibitem{threshold2}
See, for example, M.~Drees, K.~Hagiwara, and A.~Yamada,
Phys.~Rev.~D {\bf 45}, 1725 (1992);
J.~Sola, in {\em Phenomenological Aspects of Supersymmetry},
eds. W.~Hollik {\em et al.} (Springer, Berlin, 1992), p.~187;
H.~E.~Haber, in {\em Recent Directions in Particle Physics Theory},
eds. J.~Harvey and J.~Polchinski (World Scientific, Singapore, 1993);
P.~Chankowski, Z.~Pluciennik, and S.~Pokorski, 
Nucl.~Phys.~{\bf B439}, 23 (1995);
W.~Hollik, Acta Phys.~Polon.~B {\bf 27}, 3685 (1996); 
J.~Bagger, K.~Matchev, D.~Pierce, and R.-J.~Zhang,
Nucl.~Phys.~{\bf B491}, 3 (1997).

\bibitem{alphas}
I.~Antoniadis, J.~Ellis, and D.~V.~Nanopoulos, Phys.~Lett.~B 
{\bf 262}, 109 (1991);
L.~Clavelli, Phys.~Rev.~D {\bf 46}, 2112 (1992);
L.~Clavelli, P.~W.~Coulter, and K.-J.~Yuan, Phys.~Rev.~D {\bf 47}, 
1973 (1993);
M.~Jezabek and J.~H.~Kuhn, Phys.~Lett.~B {\bf 301}, 121 (1993).

\bibitem{barger}
V.~Barger, M.~S.~Berger, and R.~J.~N.~Phillips, Phys.~Lett.~B 
{\bf 382}, 178 (1996);
P.~Kraus and F.~Wilczek, Phys.~Lett.~B {\bf 382}, 262 (1996);
J.~Ellis and D.~A.~Ross, Phys.~Lett.~B {\bf 382}, 187 (1996).

\bibitem{susyrho}
R.~Barbieri and L.~Maiani, Nucl.~Phys.~{\bf B224}, 32 (1983). 
 
\bibitem{Rp}
G.~Farrar and P.~Fayet, Phys.~Lett.~B {\bf 76}, 575 (1978).

\bibitem{higgs}
S.~P.~Li and M.~Sher, Phys.~Lett.~{\bf 140B}, 339 (1984);
H.~E.~Haber and R.~Hempfling, Phys.~Rev.~Lett.~{\bf 66}, 1815 (1991);
Y.~Okada, M.~Yamaguchi, and T.~Yanagida, Prog.~Theor.~Phys.~{\bf 85}, 1
(1991);
J.~Ellis, G.~Ridolfi, and F.~Zwirner, Phys.~Lett.~B {\bf 257}, 83
(1991); 
{\em ibid.}, {\bf 262}, 477 (1991). 

\bibitem{LR}
L.~Randall, talk presented at Supersymmetry 97, Fifth International
Conference on Supersymmetries in Physics, The University of
Pennsylvania, Philadelphia, May 27--31, 1997.

\bibitem{custodial}
P.~Sikivie, L.~Susskind, M.~Voloshin, and V.~Zakharov,
Nucl.~Phys.~{\bf B173}, 189 (1980).

\bibitem{FaMa}
G.~R.~Farrar and A.~Masiero, RU--94--38, hep-ph/9410401.
Also, in the models of Ref.~\cite{gm} there are no tree-level
gaugino masses, and the parameters $\widetilde{T}_i$ defined 
below are measures of supersymmetry breaking in the messenger sector.

\bibitem{chankowski}
P. H. Chankowski, Phys.~Rev.~D {\bf 41}, 2877 (1990). 

\bibitem{erler}
J.~Erler, J.~L.~Feng, and N.~Polonsky, Phys.~Rev.~Lett. {\bf 78}, 3063
(1997).

\bibitem{ACFH} 
N.~Arkani-Hamed, H.-C.~Cheng, J.~L.~Feng, and L.~J.~Hall,
Phys.~Rev.~Lett. {\bf 77}, 1937 (1996).

\bibitem{farrar}
For a review, see G.~Farrar, talk presented at Supersymmetry 97, 
Fifth International Conference on Supersymmetries in Physics,
The University of Pennsylvania, Philadelphia, May 27--31, 1997.

\bibitem{nelson}
A.~E.~Nelson, talk presented at Supersymmetry 97, 
Fifth International Conference on Supersymmetries in Physics,
The University of Pennsylvania, Philadelphia, May 27--31, 1997. 

\bibitem{otherwork}
E.~Katz, L.~Randall, and S.~Su, MIT--CTP--2646, in preparation;
D.~Pierce, M.~Nojiri, and Y.~Yamada, SLAC--PUB--7558, in preparation;
D.~Pierce and S.~Thomas, SLAC--PUB--7474, SU--ITP 97--24, in
preparation.

\end{thebibliography}
\end{document}